\documentclass[conference,draftclsnofoot,onecolumn]{IEEEtran}
\IEEEoverridecommandlockouts
\pdfoutput=1
\usepackage{cite}
\usepackage{amsmath,amssymb,amsfonts}
\usepackage{algorithmic}
\usepackage{graphicx}
\usepackage{textcomp}
\usepackage{xcolor}
\usepackage{comment}
\def\BibTeX{{\rm B\kern-.05em{\sc i\kern-.025em b}\kern-.08em
    T\kern-.1667em\lower.7ex\hbox{E}\kern-.125emX}}
\begin{document}

\title{Environment-aware UAV Communications: CKM Construction and Predictive Beamforming
}

\author{\IEEEauthorblockN{Shiqi Zeng, Xiaoli Xu and Yong Zeng}
}

\maketitle

\section{Introduction}
\subsection{Related Work}
\subsection{Our Contributions}

\section{System Model}
As shown in Fig. 1, we consider a mmWave cellular-connected UAV system in a real urban environment, where the ground BS communicates with a single-antenna UAV and simultaneously tracks its state with the help of LPM. 
Without loss of generality, we assume that the space of interest is a cube, which can be defined by $[q_x^L,q_x^U]\times[q_y^L,q_y^U]\times[q_z^L,q_z^U]$ , with the subscripts $L$ and $U$ respectively denoting the lower and upper bounds of the 3D coordinate of the space. The LPM, denoted by $\mathcal{M}_{\text{LoS}}$, is a database stored at the BS, which returns the probability of LoS link between the BS and the potential UAV location $\mathbf{q}$. The BS is equipped with $M_t \times N_t$ transmit antennas and $M_r \times N_r$ receive antennas, where the notations $M$ and $N$ represent the row and column number of each antenna array, respectively. 
\begin{figure}[htb]
	\centering
	\includegraphics[width=0.6\textwidth,trim=0 8 35 26,clip]{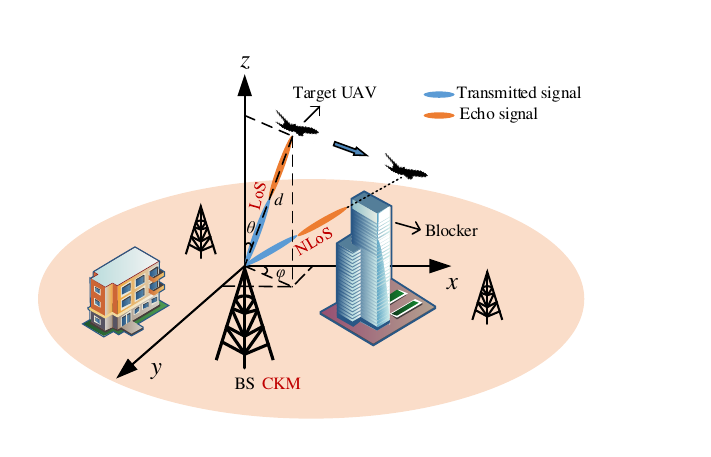}
	\caption{System model.}
	\end{figure}

\subsection{LPM Construction Model}
The LoS condition between the UAV and BS is dependent on the UAV position and environment. Divide the region of interest evenly into $M$ grids, with $\mathbf{q}_m\in \mathbb{R}^{3\times1}$ denoting the center coordinate of the $m$th cell, $1\leq m \leq M$. In this paper, we propose to utilize LPM as the prior information to boost the accuracy of link state identification. Specifically, given the BS location $\mathbf{q}_{\text{BS}}\in \mathbb{R}^{3\times1}$ and some offline supplementary information $\mathcal{W}$, LPM is a database that provides a mapping from specific receiver position $\mathbf{q}_m$ to its LoS probability $P_{\text{LoS}}(\mathbf{q}_m)$. Mathematically, LPM $\mathcal{M}_{\text{LoS}}$ can be expressed by 
\begin{equation}
\begin{aligned}
\mathcal{M}_{\text{LoS}}:\{\mathbf{q}_{\text{BS}},\mathbf{q}_m,\mathcal{W}\}\rightarrow P_{\text{LoS}}(\mathbf{q}_m)\in [0,1]\label{LPM model}.
\end{aligned}
\end{equation}
Note that the LoS probability of each position in the $m$th cell is uniformly represented by $P_{\text{LoS}}(\mathbf{q}_m)$.

Correspondingly, the NLoS probability for the position $\mathbf{q}_m$ is obtained as
\begin{equation}
\begin{aligned}
 P_{\text{NLoS}}(\mathbf{q}_m)=1-P_{\text{LoS}}(\mathbf{q}_m).\\
\end{aligned}
\end{equation}
   
\subsection{UAV State Evolution}
In this paper, we consider the constant velocity evolution model in the Cartesian coordinate system. Denote the target UAV state vector at the $n$th slot by $\mathbf{x}_n={[\mathbf{q}_n^{T},\mathbf{v}_n^{T}]^T}$, with position $\mathbf{q}_n={[{q}_{xn},{q}_{yn},{q}_{zn}]}^T$ and velocity $\mathbf{v}_n={[{v}_{xn},{v}_{yn},{v}_{zn}]}^T$. Then we have  
\begin{equation}
\begin{aligned} 
\mathbf{q}_{n+1}=\mathbf{q}_n+\mathbf{v}_n\Delta T+\mathbf{n}_d,\mathbf{v}_{n+1}=\mathbf{v}_n+\mathbf{n}_v\label{state evolution equation},
\end{aligned}
\end{equation}
\begin{equation}
\begin{aligned} 
\mathbf{q}_L\preceq \mathbf{q}_n\preceq \mathbf{q}_U, \forall n\in [1,L]\label{UAV motion limitation},
\end{aligned}
\end{equation}
where $\mathbf{n}_d, \mathbf{n}_v\in \mathbb{R}^{3\times 1}$ are Gaussian process noise, $\mathbf{q}_L=[q_x^L;q_y^L;q_z^L]$ and $\mathbf{q}_U=[q_x^U;q_y^U;q_z^U]$, with $\preceq$ denoting the element-wise inequality.

Further, the model developed in \eqref{state evolution equation} can be recast in compact form as 
\begin{equation}
\begin{aligned} 
\mathbf{x}_{n+1}=\mathbf{G}\mathbf{x}_n+\mathbf{n}_s\label{UAV motion model},
\end{aligned}
\end{equation}
where $\mathbf{n}_s$$=$${[\mathbf{n}_d^T,\mathbf{n}_v^T]^T}$ is the state noise vector with covariance matrix $\mathbf{Q}_s=\mathrm{diag}(\sigma_d^2\mathbf{I}_3^T,\sigma_v^2\mathbf{I}_3^T)$, and $\mathbf{G}$ represents the state evolution matrix.

\subsection{Sensing Model}
The BS is equipped with $M_t \times N_t$ transmit antennas and $M_r \times N_r$ receive antennas, where the notations $M$ and $N$ represent the row and column number of each antenna array, respectively. Consider a time duration of $T$, which is discretized into $L$ small time slots with a length of $\Delta T$. Define the downlink signal transmitted by BS at the $n$th slot and time $t$ as $s_n(t), t\in [(n-1)\Delta T,n\Delta T), 1\leq n \leq L$. The corresponding reflected signal is denoted by $\mathbf{r}_n(t)\in \mathbb{C}^{M_rN_r\times 1}$, and we use $d_n$, $\varphi_n$, and $\theta_n$ to represent the distance, azimuth angle, and elevation angle between the reflector and the BS at the $n$th slot, respectively. 

When the LoS link exists, the BS is expected to receive the reflected signal from the UAV, which is given by
\begin{equation}
\begin{aligned}
\mathbf{r}_n^o(t)=&{{\alpha_{n} ^{o}}} \sqrt {p_{n}M_rN_rM_tN_t}   {e^{j{2\pi \mu _{n}^{o}}t}}{\mathbf {a}_r}\left ({{{\varphi_n,\theta _{n}}} }\right)\\&{{\mathbf {a}_t}^{H}}\left ({{{\varphi_n,\theta _{n}}} }\right){{\mathbf {f}_n}}  {s}_{n}(t - \tau _{n}^{o}) + \mathbf{z}_{n}^{r}\left ({t }\right), \\
\end{aligned}
\end{equation}
where the superscript $o$ indicates that parameters are associated with the target UAV. $\alpha_n^o$ denotes the reflection coefficient, which is mainly determined by the UAV's radar cross section (RCS). $p_n$ is the transmit power at the $n$th slot. $\mathbf{f}_n\in\mathbb{C}^{M_tN_t\times1}$ represents the transmitting beamforming vector. Furthermore, $\mu_n^o=2f_cv_{n,r}^o/c$ is the Doppler frequency, where $v_{n,r}^o$ denotes the radial velocity of the UAV, $f_c$ and $c$ represent carrier frequency and light speed, respectively. $\tau_n^o=2d_n^o/c$ is the time delay, where $d_n^o$ is the distance between the BS and UAV. $\mathbf{z}_n^r(t)\sim \mathcal{CN}(0,\sigma_r^2\mathbf{I}_{M_rN_r})$ is the complex additive white Gaussian noise. ${\mathbf {a}_t}\left ({{{\varphi,\theta }} }\right)$ and ${\mathbf {a}_r}\left ({{{\varphi,\theta }} }\right)$ represent the transmit and receive steering vectors of the antenna array of the BS, respectively, which are given by
\begin{equation}
\begin{aligned}
\mathbf{a}_{t}(\varphi,\theta)=\frac{1}{\sqrt{M_tN_t}}[&1,...,e^{-j\pi \sin \theta[(m-1)\cos\varphi+(n-1)\sin \varphi]},\\&...,e^{-j\pi \sin \theta[(M_t-1)\cos\varphi+(N_t-1)\sin \varphi]}]^T,\\
\mathbf{a}_{r}(\varphi,\theta)=\frac{1}{\sqrt{M_rN_r}}[&1,...,e^{-j\pi \sin \theta[(m-1)\cos\varphi+(n-1)\sin \varphi]},\\&...,e^{-j\pi \sin \theta[(M_r-1)\cos\varphi+(N_r-1)\sin \varphi]}]^T,\\
\end{aligned}
\end{equation}
where the antenna element spacing is set as half wavelength.

If the LoS link between the BS and UAV is blocked, the echo signal received at the BS is reflected by the blocker, which can be expressed as
\begin{equation}
\begin{aligned}
\mathbf{r}_n^b(t)=&{{\alpha_{n} ^{b}}} \sqrt {p_{n}M_rN_rM_tN_t}   {e^{j{2\pi \mu _{n}^{b}}t}}{\mathbf {a}_r}\left ({{{\varphi_n,\theta _{n}}} }\right)\\&{{\mathbf {a}_t}^{H}}\left ({{{\varphi_n,\theta _{n}}} }\right){{\mathbf {f}_n}}  {s}_{n}(t - \tau _{n}^{b}) + \mathbf{z}_{n}^{r}\left ({t }\right), \\
\end{aligned}
\end{equation}
where parameters related to the blocker are tagged with the superscript $b$. As compared with (1), the echo signal reflected by the blocker may have different reflection coefficients, Doppler frequencies, and time delays. 

\subsection{Communication Model}
The downlink communication and sensing share the same transmitted signal. The received signal by the UAV when LoS link exists is 
\begin{equation}
\begin{aligned}
{r}_n^{c,\text{LoS}}=\kappa_n^{\text{LoS}}\sqrt{p_nM_tN_t}{\mathbf{a}}_t^{H}(\varphi_n,\theta_n)\mathbf{f}_ns_n(t-\tau_n^{c,\text{LoS}})+z_n^{c}(t),
\end{aligned}
\end{equation}
where communication-related parameters are marked with superscript $c$, $\tau_n^{c,\text{LoS}}$ is the time delay, $z_n^c(t)\sim \mathcal{CN}(0,\sigma_c^2)$ denotes the Gaussian noise, and $\kappa_n^{\text{LoS}}$ is the channel coefficient, which is modelled as
\begin{equation}
\begin{aligned}
\kappa_n^{\text{LoS}}=\kappa_{\text{ref}}e^{j\frac{2\pi f_c}{c}d_n^o}\big/ { d_n^o},\label{kappa}
\end{aligned}
\end{equation}
where $\kappa_{\text{ref}}$ is the reference path loss at the distance of 1m.

Since that the transmitted signal $s_n(t)$ has a unit power, the received signal-to-noise ratio (SNR) for the UAV at the $n$th slot is given by
\begin{equation}
\begin{aligned}
\text{SNR}_n^{\text{LoS}}=\frac{p_nM_tN_t|\kappa_n^{\text{LoS}}|^2|\mathbf{a}_t^H(\varphi_n,\theta_n)\mathbf{f}_n|^2}{\sigma_c^2}.\label{LoS SNR}
\end{aligned}
\end{equation}

When the LoS link is blocked, the received SNR will degrade significantly, which is represented by $\text{SNR}_n^{\text{NLoS}}$. 
In this case, the mmWave link handover may be performed to mitigate the adversary effect from the blockage. Specifically, once NLoS link state is identified at the $n$th slot, the target UAV may be connected to another BS in the cellular network immediately to avoid communication outage. With the notation of $B_k$, $k\in [1,2,...,K]$, we assume that there are $K$ BSs in the area of interest. The UAV tends to be associated with the nearest BS, without loss of generality, we set $B_1$ as the BS that always keeps the nearest distance from the target UAV during the observation period. NLoS identification triggers handover, and then a candidate $B_k$, $k\neq1$ is expected to be selected according to specific handover criterion. Correspondingly, we use $\text{SNR}_n^{\text{h-o}}$ to denote the UAV received SNR, more details will be discussed in section V.


\section{Environment-aware UAV communications }
To establish reliable communication links without extensive beam training, the BS may track the UAV trajectory and adopt the predictive beamforming\cite{9171304}. Conventionally, the beam alignment is performed by \textit{beam training}, which periodically transmits and receives pilots at all the possible beams, and then searches for the beam pair that gives the strongest path gain\cite{7914742}. Obviously, such a process will inevitably cause considerable latency and overhead brought by pilots and feedbacks between the transmitter and receiver. To this end, The technique of \textit{predictive beamforming} is expected to efficiently reduce the overhead incurred in beam training. Specifically, when the receiver's trajectory is well-tracked, the BS predicts the next position of the receiver, and then transmits a formulated beam towards the predicted direction. Moreover, in this work UAV tracking and communication will share the same mmWave signal to avoid extra overhead of sensing devices.

On the other hand, due to the high susceptibility of mmWave signal to the blockage, we also need to consider the impact of NLoS link between the UAV and BS in the tracking process. Thus, we propose to construct LPM to enable LoS/NLoS identification at the BS side in real-time. Instructed by the link state identification result, UAV tracking, predictive beamforming and UAV association will be performed sequentially.      
Overall, to accomplish both sensing and communication tasks reliably in the mmWave cellular-connected UAV system, we propose a LPM-assisted LoS identification and predicitive beamforming scheme, and the overview is shown in Fig. 2. 
\begin{figure}[htb]
	\centering
	\includegraphics[width=0.7\textwidth]{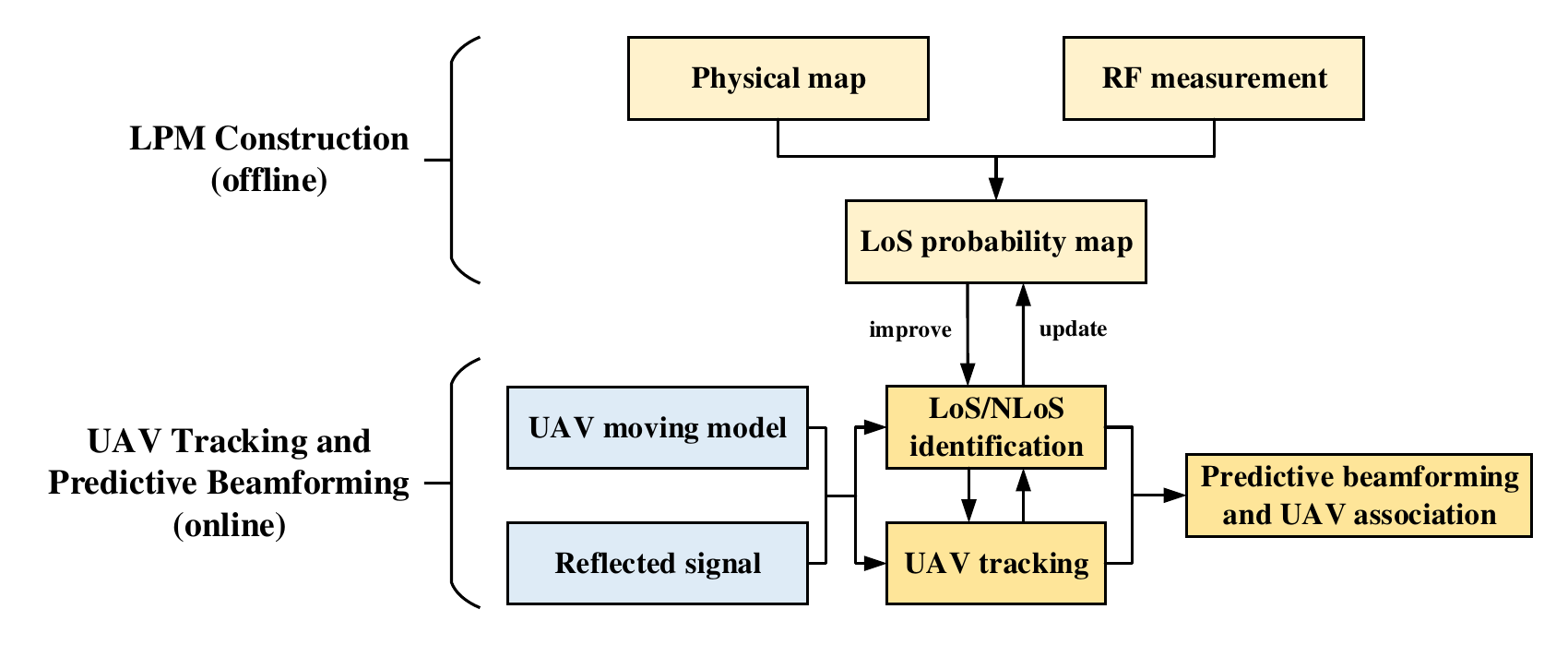}
	\caption{The proposed environment-aware UAV communications.}
	\end{figure}

In the light of the above discussion, the proposed scheme can be be implemented in the following two phases:

Phase 1 mainly explores an offline CKM construction method. Considering the sensitivity of millimeter wave signal to LoS link blockage, this work chooses to construct LoS probability map (LPM) \cite{zeng2023ckm} to help the system to improve the accuracy of UAV tracking and promote the system's communication performance. Specifically, when the location of the ground BS is known, LPM can give the LoS probability between the BS and any interested user location. The prior LoS probability information provided by LPM will assist BS in performing real-time link state identification during the UAV tracking process. On the other hand, LPM can be incorporated in BS association strategy selection for alleviating communication outages caused by the NLoS link. The offline information required to build LPM mainly includes the Two Dimensional Physical Map and Radio-Frequency Measurements. The specific construction method will be discussed in Section IV.

Phase 2 focus on the online UAV tracking and predictive beamforming. During this process, the UAV is connected to the cellular network as an aerial user, and the ground BS will transmit downlink millimeter wave ISAC signals to communicate with UAV and then track UAV's location through the echoes. Note that the LoS link between the UAV and BS may be obstructed by buildings during low altitude flight, so it is necessary to perform the link state identification during target tracking. In this regard, this work proposes to implement UAV trajectory tracking based on UAV motion model and reflected signals, while taking the NLoS link impact into consideration. To mitigate communication outage and boost the communication rate, this work also investigates corresponding communication strategies for different results of  link state identification.  Specifically,  In the case of LoS, the BS will perform predictive beamforming based on the predicted trajectory of the UAV to avoid the overhead of beam training. While in the NLoS case, the UAV will immediately associate with another BS based on the prior information provided by LPM to bypass communication interrupt caused by obstacles. The specific communication scheme will be discussed in Section V.

\section{Physical Map-Assisted LPM Construction}
\subsection{Prior LPM Construction }
\subsection{LPM Improvement with RF Measurement}

\section{LPM-Assisted UAVTracking and Predictive Beamforming}
\subsection{UAV Tracking}
\subsection{LoS Identification}
\subsection{Predictive beamforming and BS associateion}

\section{Numerical Results}
\subsection{LPM Evaluation}
\subsection{Communication Performance}
\section{Conclusion}

\bibliographystyle{ieeetr}
\bibliography{reference}

\end{document}